%% file: root.tex
\documentclass[letterpaper, 10 pt, conference]{ieeeconf}  

\IEEEoverridecommandlockouts                              

\usepackage{graphics} 
\usepackage{epsfig} 
\usepackage{mathptmx} 
\usepackage{times} 
\usepackage{amsmath} 
\usepackage{amssymb}  
\usepackage{arydshln} 
\usepackage{hyperref}

\title{\LARGE \bf
Improved Decoding of Attentional Selection in Multi-Talker Environments with Self-Supervised Learned Speech Representation
}

\author{Cong Han$^{1}$, Vishal Choudhari$^{1}$, Yinghao Aaron Li$^{1}$ and Nima Mesgarani$^{1}$
\thanks{$^{1}$Department of Electrical Engineering, Columbia University, New York, NY. Cong Han: {\tt\small ch3212@columbia.edu}; Nima Meagarani: {\tt\small nima@ee.columbia.edu}}%
}

\begin{document}

\maketitle
\thispagestyle{empty}
\pagestyle{empty}

\begin{abstract}
\input{abstract.tex}

\end{abstract}

\section{INTRODUCTION}
\input{introduction.tex}

\section{Related work}
\input{background.tex}

\section{Materials and Method}
\input{method.tex}

\section{Results}
\input{results.tex}

\section{CONCLUSIONS}
\input{conclusion.tex}

\addtolength{\textheight}{-12cm}   




\section*{ACKNOWLEDGMENT}

{This work was funded by the national institute of health (NIH-NIDCD) and a grant from Marie-Josée and Henry R. Kravis. We used ChatGPT to improve the clarity and readability of writing.}


\bibliographystyle{IEEEtran}
{\bibliography{refs}}

\end{document}

%% file: abstract.tex
Auditory attention decoding (AAD) is a technique used to identify and amplify the talker that a listener is focused on in a noisy environment. This is done by comparing the listener's brainwaves to a representation of all the sound sources to find the closest match. The representation is typically the waveform or spectrogram of the sounds. The effectiveness of these representations for AAD is uncertain. In this study, we examined the use of self-supervised learned speech representation in improving the accuracy and speed of AAD. We recorded the brain activity of three subjects using invasive electrocorticography (ECoG) as they listened to two conversations and focused on one. We used WavLM to extract a latent representation of each talker and trained a spatiotemporal filter to map brain activity to intermediate representations of speech. During the evaluation, the reconstructed representation is compared to each speaker's representation to determine the target speaker. Our results indicate that speech representation from WavLM provides better decoding accuracy and speed than the speech envelope and spectrogram. Our findings demonstrate the advantages of self-supervised learned speech representation for auditory attention decoding and pave the way for developing brain-controlled hearable technologies.


%% file: introduction.tex
Hearing-impaired listeners often struggle with understanding speech in noisy or crowded environments. A brain-controlled assistive hearing device can help by selectively amplifying the speech of the talker the listener is paying attention to \cite{han2019speaker}. This device relies on a technique called auditory attention decoding (AAD), which uses brain activity to determine which talker the listener is attending to. There are various ways of acquiring brain activity for AAD, such as magnetoencephalography (MEG) \cite{ding2012emergence}, electroencephalography (EEG) \cite{o2015attentional}, and electrocorticography (ECoG) \cite{mesgarani2012selective}. Most AAD algorithms reconstruct a representation of speech from the brain activity and compare it to all the talkers in the environment. The talker with the highest correlation is considered the attended talker. Typically, the representation of sound that is used in AAD is either the envelope or spectrogram. However, it is not clear if either of these are optimal for neural decoding. A good representation of sound for AAD should be easily reconstructed from brain activity. In particular, learning a complex nonlinear mapping from brain signals to the representation can be challenging due to the limitations in recording the brain signals. Additionally, a good representation should have a stronger correlation with the attended speaker than with unattended speakers.

Self-supervised representation learning (SSL) for speech has been successfully applied in many applications \cite{baevski2020wav2vec,hsu2021hubert,chen2022wavlm}. SSL learns representations through designed pretext tasks, where the input and learning targets are derived from the input signal itself. Because of this, SSL can be easily scaled up with a large amount of unlabeled speech data. The self-supervised learned representations are often used as input features for downstream tasks to reduce the need for a large amount of labeled training data and improve task performance. Studies have shown that the learned speech representations can improve various downstream tasks such as speech recognition, speaker identification, and intent classification \cite{yang2021superb}. A recent study indicates that the functional hierarchy of latent layers of a self-supervised speech model aligns well with the cortical hierarchy of speech processing \cite{millet2022toward}. Additionally, the learned representations have been found to be more effective at predicting cortical responses to speech than hand-engineered acoustic features \cite{vaidya2022self}, motivating the idea that they may be related to attention and superior to traditional acoustic features used in the AAD task.

In this study, we used self-supervised learned speech representations to improve the neural decoding of attentional selection. We used an intermediate layer of the pre-trained WavLM model \cite{chen2022wavlm} as the reconstruction target from the brain signals instead of using traditional speech envelope and spectrogram features. Our experimental results show a significant improvement in decoding accuracy when using these learned speech representations. Additionally, we modified the WavLM model to a causal configuration to test its potential for real-time implementation and found that it still outperforms speech envelope and spectrogram features, indicating that transformer-based self-supervised representations are superior candidates for brain-controlled hearable devices. 

\begin{figure*}[!t]
  \centering
  \includegraphics[width=2\columnwidth]{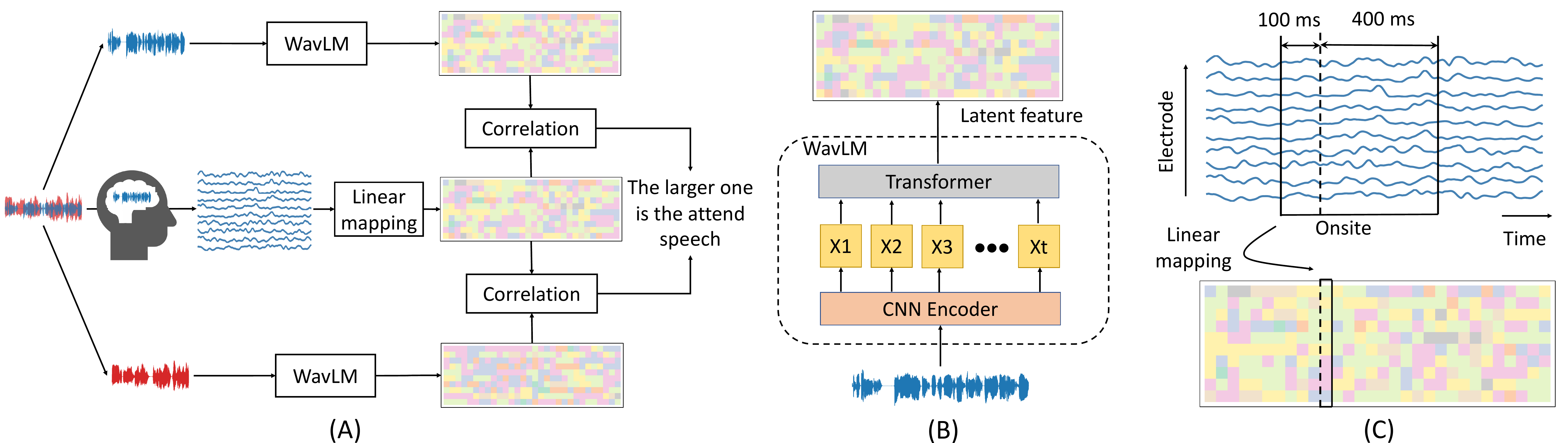}
  \caption{Diagram showing the process of auditory attention decoding. (A) Neural activity is monitored while a subject listens to a mixture of two talkers and focuses on one of them. WavLM extracts the representation of individual talkers obtained through a speech separation model. The predicted representation from the neural activity is compared to the representation of individual talkers to determine the most similar talker. (B) WavLM consists of a CNN encoder and transformer, producing layers of features. One intermediate layer is used as the speech representation. (C) Linear spatiotemporal filters map the neural activity, with time-lags ranging from $-400$ ms to 100 ms, to the learned representation.
  }
  \label{diagram}
\end{figure*}

%% file: background.tex
\subsection{Stimulus reconstruction for AAD}
Recent research in neuroscience has revealed that the reconstructed spectrogram from neural responses in the human auditory cortex is most similar to the spectrogram of the talker the subject is focusing on \cite{mesgarani2012selective}. To determine the attended talker, the reconstructed spectrogram from neural signals is  compared to the spectrograms of individual talkers. Linear methods are commonly used for this mapping, where a linear time-invariant filter is applied to the set of electrodes to predict the attended speech spectrogram. The linear filter is optimized by minimizing the mean-squared error between the reconstructed and actual spectrogram. The speech envelope can also be used as a reconstruction target, particularly when the neural data has a low signal-to-noise ratio (SNR), such as in the case of EEG data. This study aims to investigate whether reconstructing the self-supervised learned speech representations from neural activity can improve the accuracy of decoding the attended speaker.

\subsection{WavLM}
Several self-supervised speech representation learning approaches have been proposed recently, with wav2vec 2.0 \cite{baevski2020wav2vec} and HuBERT \cite{hsu2021hubert} being two of the most well-known. Both models have a similar architecture but differ in their pre-training strategies. Wav2vec 2.0 uses a contrastive loss to differentiate between positive and negative samples, while HuBERT uses an offline clustering approach to assign labels to speech units, and then trains the model through a BERT-like masked speech prediction task. This forces the model to learn both acoustic and language features from unlabeled speech data. WavLM, a variant of HuBERT, adds a speech denoising task during pre-training to improve its ability to handle non-ASR tasks such as speech diarization and separation. WavLM Large, trained on 94k hours of diverse speech data, outperforms previous self-supervised speech models on SUPERB \cite{yang2021superb}, demonstrating its high capacity to model speech speaker, content, and semantics. Given its strong performance, we selected WavLM Large for this study, as it has the best potential to improve attended-or-unattended talker classification accuracy. Decoding attentional selection is therefore regarded as a downstream task of WavLM.

%% file: method.tex
\subsection{Neural Data Acquisition and Preprocessing}
We used invasive intracranial electroencephalography (iEEG) to measure neural activity from three neurosurgical patients undergoing treatment for epilepsy. Two patients (Subjects 1 and 2) had stereo-electroencephalography (sEEG) depth as well as subdural electrocorticography (ECoG) grid electrodes implanted over the left hemispheres of their brains. The third patient (Subject 3) only had sEEG depth electrodes implanted over their left brain hemisphere. All subjects had self-reported normal hearing and consented to participate in the experiment.

Each subject participated in 28 trials. In each trial, the subjects listened to a mixture of two conversations amidst a background noise. Two native American English talkers took turns in each conversation. The two conversations were concurrent, independent and moving in the frontal half of the horizontal plane of the subject. Both conversations were of equal power. The background noise was either 
street noise or speech babble noise. The subjects were cued to focus their attention on one conversation and to ignore the other. To ensure that the subjects were engaged in the task, we inserted repeated words (RWs) in both conversation streams and asked the subjects to press a button whenever they heard a RW in the conversation being followed. The subjects were able to detect the RWs in the cued conversation with the following accuracies: S1 - 73.2\%, S2 - 74.8\%, S3 - 67.3\%.

We pre-processed the neural data to extract its high gamma (70 - 150 Hz) envelope by following the same procedure described in ``Data Preprocessing and Hardware" section in \cite{han2019speaker}. Electrodes were chosen if they were significantly more responsive to speech than to silence. The Subjects S1, S2 and S3 had 34, 42 and 17 speech-responsive electrodes, respectively. The sampling rate of the neural data was 100Hz.

\subsection{Extraction of Speech representations}
The WavLM Large model \footnote{The pre-trained model can be found at {\url{https://github.com/microsoft/unilm/tree/master/wavlm}}} is composed of a convolutional encoder and 24 transformer layers. The convolutional encoder converts a waveform sampled at 16 kHz to a feature sequence at a 50 Hz framerate (one frame every 20ms), with each frame encoding about 25 ms of the waveform. Each transformer layer has an embedding dimension of 1024 and 12 self-attention heads. 

We used WavLM to extract the latent representation $X \in \mathbb{R}^{1024*T}$ of speech waveforms from the i-th layer, where T is the number of time frames. We upsampled X to 100 Hz to match the rate of the neural data. Since our speech duration is long, we limited the attention span of each frame in the transformer layers to 6 seconds ($\sim 300$ time frames) with 3 seconds before and 3 seconds after the frame. However, WavLM is a noncausal model because each time frame attends to future time frames. For a fair comparison with the speech envelope and spectrogram, and to enable real-time AAD, we modified WavLM for a causal configuration. First, we set the attention weights of all the future frames as zero to force each frame to only attend to the past 6 seconds, and we refer to this model as ``WavLM w/ causal ATT". The transformer layer in WavLM is equipped with a convolution-based position embedding where the convolution operation has access to the future frames, which results in noncausal computation. To avoid this, we changed the noncausal convolution to a causal convolution, and the resulting model is referred to as ``WavLM w/ causal ATT \& PE". Note that we did not finetune WavLM after we modified the attention weights or positional embedding. In case of random effects that cause performance gain for WavLM, we added WavLM with random initialization as a control. 

\subsection{Representation Reconstruction for Decoding Attention}
We adopted the linear reconstruction method \cite{mesgarani2009influence}. Specifically, two subject-wise liner spatiotemporal filters $G_A$ and $G_u$ were learned to map neural activity R to the speech representations of the attended ($\hat{X}_A$) and unattended ($\hat{X}_U$) talkers,
\begin{align}
\label{eqn:eq1}
    &\hat{X}_A(n, t) = \sum_e\sum_{\tau}G_A(n, e, \tau)R(e, t-\tau) \\
    &\hat{X}_U(n, t) = \sum_e\sum_{\tau}G_U(n, e, \tau)R(e, t-\tau),
\end{align}
where $n$ is the channel index of the representation, $e$ is the neural electrode index, and $\tau$ is the time lag, ranging from $-400$ ms to 100 ms in this study. As before, the linear filters were optimized by minimizing the mean-squared errors between the reconstructed and the actual representations. 

A leave-one-out cross-validation approach was used, wherein the subject-wise filters were trained on N - 1 trials and used to reconstruct representations $\hat{X}_A, \hat{X}_U$ on the left out trial. We calculated Pearson’s correlation coefficient between the reconstructed representations $\hat{X}_A, \hat{X}_U$ and the representations of two talkers  $X_{sp1}, X_{sp2}$. The correlation coefficient is estimated across a window of seconds, which is referred to as the decoding window duration. We used sliding window of 0.5 s, 1 s, 2 s, 4 s, and 8 s, respectively, throughout the trial duration. We defined an attentional modulation index (AMI) as,
\begin{align}
\begin{split}
    AMI = c&orr(\hat{X}_A, X_{sp1}) - corr(\hat{X}_A, X_{sp2}) \\
    &+ corr(\hat{X}_U, X_{sp2}) - corr(\hat{X}_U, X_{sp1}).
\end{split}
\end{align} 
A positive value of this index suggests that speaker 1 is the attended speaker, and a negative value votes speaker 2 to be the attended speaker for this window. Decoding accuracy is defined as the percentage of windows that were correctly classified.

%% file: results.tex
\begin{figure}[t]
  \centering
  \includegraphics[width=1\columnwidth]{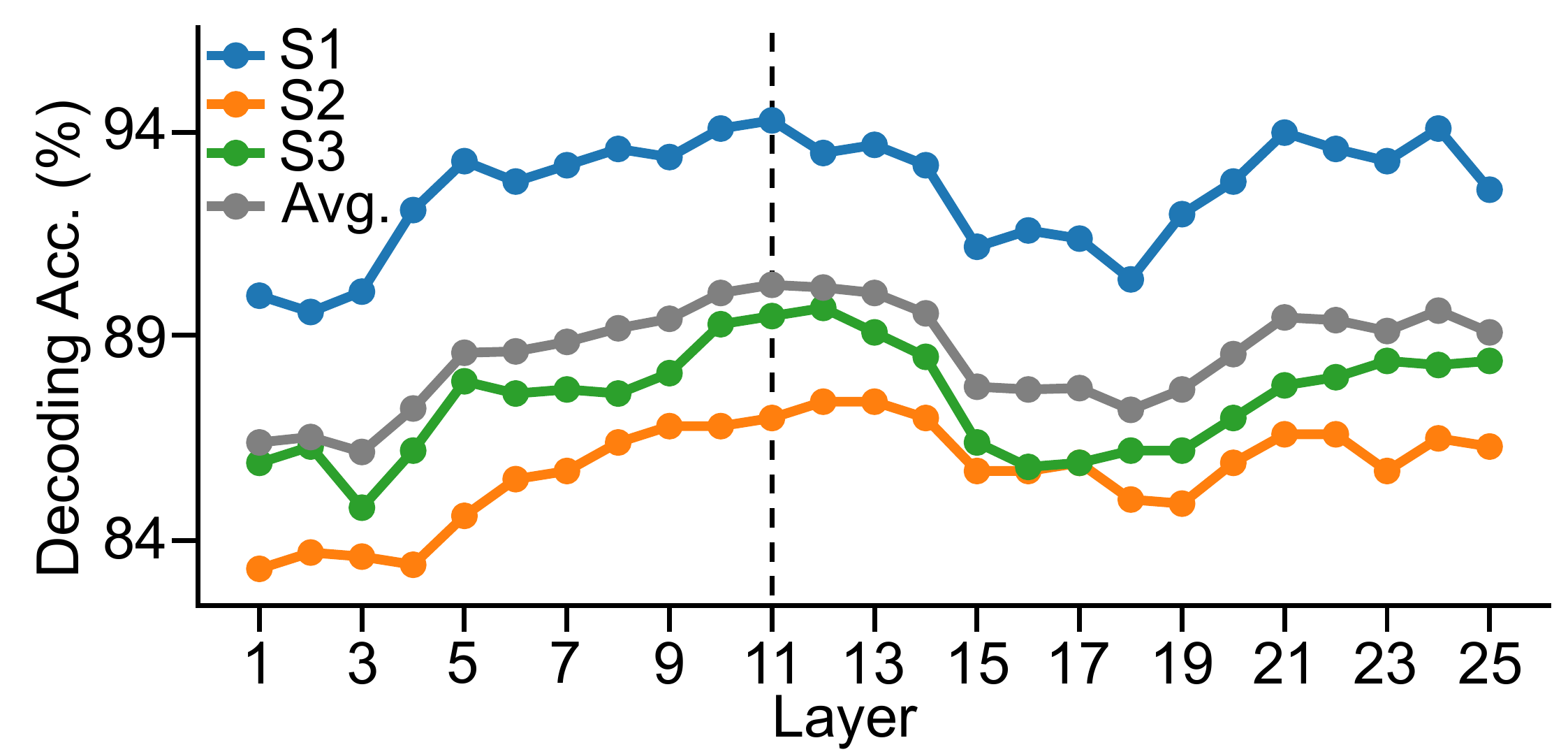}
  \caption{Accuracy of attention decoding using representations from each layer of WavLM with a 4-second decoding window. The 11-th layer shows the best average performance across subjects.}
  \label{fig2}
\end{figure}

\subsection{Decoding Accuracy for Each Layer of WavLM}
Fig. ~\ref{fig2} shows the effect of different layer representations from WavLM on decoding accuracy. The accuracy improves as the layer depth increases, then slightly decreases before climbing again. The 11-th layer produces the best performance on average for the three subjects. The first layer, which is the output of the CNN encoder, extracts local features ($\sim 25$ ms) from speech, resembling a spectrogram. The subsequent layers contain semantic information with more context. A recent layer-wise analysis of wav2vec 2.0 found an acoustic-linguistic hierarchy in layer-wise representation evolution, where shallow layers encode local acoustic information, followed by phonetics, word identity, and word meaning \cite{pasad2021layer}. Therefore, Fig. ~\ref{fig2} suggests that speech's higher-level features may be better decoded from the brain to enhance attention decoding accuracy.  \cite{pasad2021layer} also noticed a reverse trend starting from the middle layer, which they attributed to the transformer layers' autoencoder-style behavior where deeper layers become closer to the input. 

\begin{table}[b]
\caption{Accuracy of Attention Decoding Using Various Features Extracted from Clean Speech (Avg. Over 3 Subjects, in \%)}
\label{t1}
\begin{center}
\begin{tabular}{|l||ccccc|}
\hline
Feature &  \multicolumn{5}{c|}{Decoding window size} \\ 
& 0.5s & 1s & 2s & 4s & 8s  \\ 
\hline
Envelope & 63.3 &71.6 &79.5 &86.0 &91.3 \\
Mel-spectrogram & 65.6 &72.3 &80.8 &88.5 &91.5 \\
WavLM &\bf{72.9} &\bf{78.7} &\bf{85.2} &\bf{90.3} &\bf{92.6} \\
{WavLM w/ causal ATT} &72.2 &78.5 &84.6 &89.4 &92.3 \\
{WavLM w/ causal ATT \& PE} &72.0 &77.9 &84.1 &89.1 &92.5 \\
\hdashline
{WavLM w/ random init.} & 62.8 & 68.4 & 74.1 & 79.8 & 87.1 \\
\hline
\end{tabular}
\end{center}
\end{table}

\begin{table}[ht]
\caption{Accuracy of Attention Decoding Using Various Features Extracted from Separated Speech (Avg. Over 3 Subjects, in \%)}
\label{t2}
\begin{center}
\begin{tabular}{|l||ccccc|}
\hline
Feature &  \multicolumn{5}{c|}{Decoding window size} \\ 
& 0.5s & 1s & 2s & 4s & 8s  \\ 
\hline
Envelope & 63.0 &70.1 &78.5 &85.2 &90.1 \\
Mel-spectrogram & 64.6 &71.1 &79.2 &86.3 &90.4 \\
WavLM & \bf{72.2} &\bf{77.8} &\bf{83.9} &\bf{88.6} &\bf{92.4} \\
{WavLM w/ causal ATT} &70.8 &76.6 &82.5 &88.1 &92.1 \\
{WavLM w/ causal ATT \& PE} &70.6 &76.1 &82.3 &88.1 &91.7 \\
\hdashline
{WavLM w/ random init.} & 62.8 & 68.4 & 74.1 & 79.8 & 84.6\\
\hline
\end{tabular}
\end{center}
\end{table}

\subsection{Decoding Accuracy for Different Reconstruction Targets}
Table ~\ref{t1} compares the results of different features used for auditory attention decoding. The acoustic features envelope  and 28-basis Mel-spectrogram are the baseline features. The envelope and Mel-spectrogram features were Z-scored before training and inference. Results show that all the features extracted from WavLM consistently outperform the baseline (paired t-test, p $<$ 0.001 for win sizes 0.5 s, 1 s, 2 s, and 4 s; p $<$ 0.05 for win size 8 s). WavLM performs especially well compared to the baseline when the decoding window size is small. The causal configuration resulted in a slight performance decrease, but it is expected that further fine-tuning can reduce this decrease. A control experiment using WavLM with random initialization shows significantly worse results than the baseline, confirming that the performance gain for WavLM was due to better representations learned through self-supervised learning, not due to its architecture or feature dimension. 

Because clean speech of individual speakers is usually unavailable, we used an automatic speech separation model \cite{han2021binaural} to separate the mixed speech. Results in Table ~\ref{t2} show a slight decrease in accuracy compared to those in Table 1 due to imperfect speech separation, but this difference is small and all features are similarly affected. Despite this, WavLM remains superior to speech envelope and Mel-spectrogram.

\subsection{Number of Principle Components}

\begin{table}[b]
\caption{Accuracy of Attention Decoding using WavLM Feature with Various Component Numbers.}
\label{t3}
\begin{center}
\begin{tabular}{|l||ccccc|}
\hline
Feature &  \multicolumn{5}{c|}{Decoding window size} \\ 
& 0.5s & 1s & 2s & 4s & 8s  \\ 
\hline
Mel-spectrogram (28 dims) & 64.6 &71.1 &79.2 &86.3 &90.4 \\
{WavLM causal (1024 dims)} &70.6 &76.1 &82.3 &88.1 &91.7 \\
{\qquad \qquad 200 PCs} &70.3 &75.7 &82.2 &88.1 &91.7 \\
{\qquad \qquad 100 PCs} &70.0 &75.5 &81.9 &87.8 &91.7 \\
{\qquad \qquad \ 50 PCs} &69.5 &75.2 &81.8 &87.4 &91.4 \\
{\qquad \qquad \ 28 PCs} &69.0 &74.8 &81.5 &87.7 &91.7 \\
\hline
\end{tabular}
\end{center}
\end{table}

We reduced the dimension of WavLM features using Principal Component Analysis (PCA). Table~\ref{t3} presents the decoding accuracy for WavLM features with varying numbers of PCA components. Although accuracy decreases slightly with fewer components, WavLM surpasses Mel-spectrogram notably.

\subsection{Dynamic Switching of Attention}

\begin{figure}[!t]
  \centering
  \includegraphics[width=0.9\columnwidth]{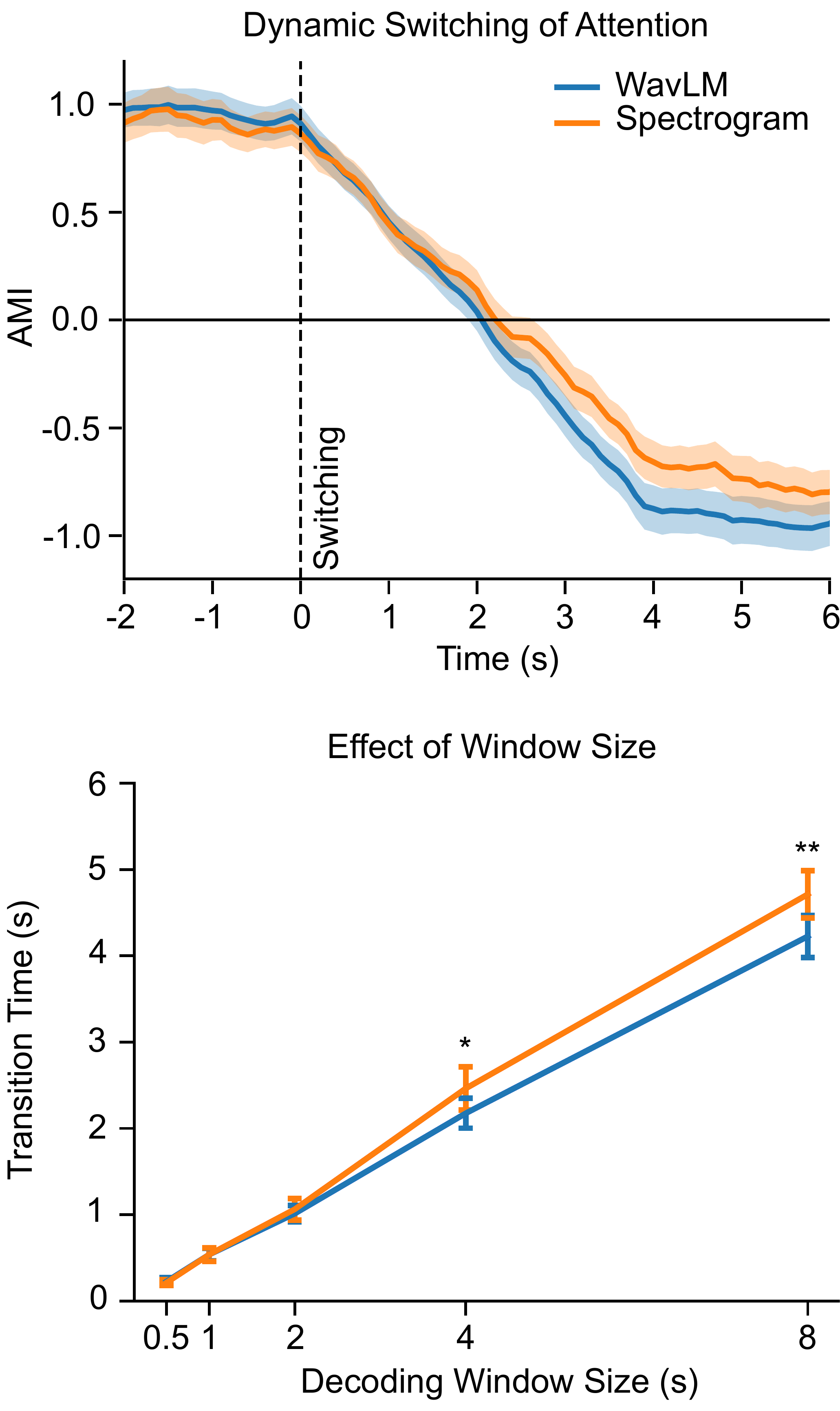}
  \caption{(1) The upper plot displays attention switching from speaker 1 to speaker 2. The dashed line represents the switch in attention. The average AMI for the three subjects is shown with a 4-second decoding window. (2) The bottom plot shows the transition time for detecting a switch which was measured as the moment when AMI crosses zero. Statistical significance is indicated by asterisks: * for $p < 0.05$ and ** for $p < 0.01$.}
  \label{figure3}
\end{figure}

We simulated dynamic attention switching by concatenating the first 10 seconds and last 10 seconds of neural responses in each trial where the subject was attending to Spk1 in the first 10 s and switched to Spk2 afterward. We calculated AMI scores for WavLM w/ causal ATT \& PE and Mel-spectrogram, respectively, using a sliding window of 4s. The upper panel of Fig.~\ref{figure3} shows the AMI scores averaged over all the subjects and trials. The averaged AMI scores were scaled between -1 and 1. WavLM and spectrogram exhibit a similar pattern but WavLM detects the switch faster. The bottom panel of Fig.~\ref{figure3} shows the average transition times for five different sliding window durations. As expected, the transition times increase for longer durations. There are no significant differences between WavLM and spectrogram for window size 2 s and below (paired t-test, $p > 0.2$); However, WavLM has a shorter transition time for window sizes 4 s ($p < 0.05$) and 8 s ($p < 0.01$).

\subsection{Comparison of Features in Predicting Neural Activity}
\begin{figure*}[!t]
  \centering
  \includegraphics[width=2\columnwidth]{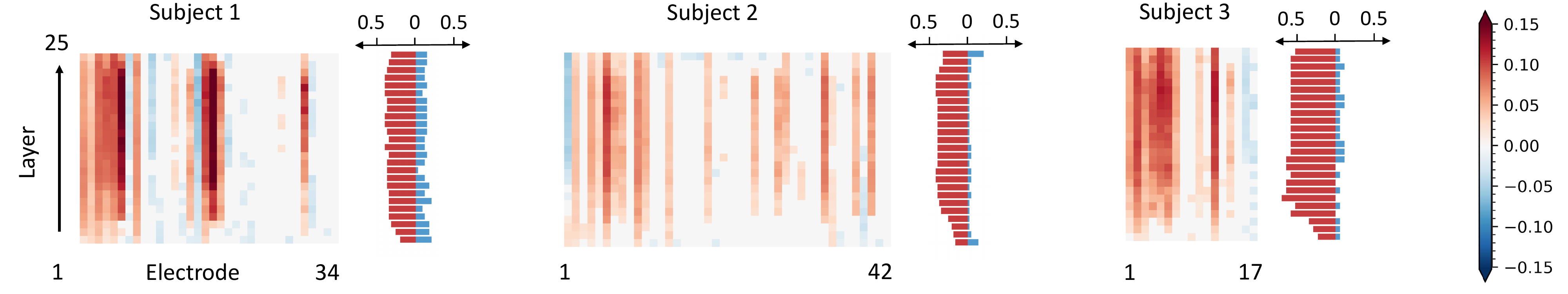}
  \caption{The improvement in r-value between the actual and predicted neural activity using WavLM features compared to using spectrogram for each layer and electrode. A positive value indicates WavLM features provide a better prediction of the neural activity at that electrode than the spectrogram, while a negative value indicates that the spectrogram is more accurate. Zeros (show in white) indicate no significant difference between the two features ($p > 0.05$). The bar plots show the proportion of electrodes that are more responsive to WavLM features (red) and to spectrogram (blue), respectively.}
  \label{heat}
\end{figure*}
To gain further insight into why SSL features provide a higher neural decoding accuracy, we used a forward model to predict the response of single neural sites from different layers of WavLM. While the stimulus reconstruction method uses a backward model, here we trained forward models, spatiotemporal filters, $G_R$, that predict neural activity based on various stimulus features, 

\begin{align}
\label{eqn:snr}
    &\hat{R}(e, t) = \sum_e\sum_{\tau}G_R(e, n, \tau)X_{A}(n, t-\tau),
\end{align}
where the time lag $\tau$ ranges from 0 to 200 ms. We measured the correlation (r-value) between the reconstructed and actual neural activity for each electrode. We assessed the improvement in r-value using each layer of WavLM compared to the Mel-spectrogram, where we utilized the first 100 PCs of WavLM features. If the results of a paired t-test showed no statistical difference between using WavLM features and the Mel-spectrogram (with a $p > 0.05$), the improvement value was set to zero. Additionally, we calculated the percentage of electrodes that were more accurately predicted using spectrogram and the percentage of electrodes that were more accurately predicted using WavLM features for each layer.

The results for the three subjects are presented in Fig.~\ref{heat}. The middle layers of WavLM generally provided better predictions compared to the shallow layers and the deepest layers. Although some electrodes were more accurately predicted using the acoustic spectrogram (shown in blue), a larger proportion of electrodes were better predicted using WavLM features (shown in red). The results in Fig.~\ref{heat} indicate that different regions of the auditory cortex encode different levels of speech information. This inspires combining different layers of WavLM features to further improve AAD accuracy, which will be the focus of future work.

%% file: conclusion.tex
This study examined the use of self-supervised speech representations to enhance attentional decoding in multi-talker situations. Results showed that substituting traditional speech features with latent features from WavLM resulted in improved attention decoding accuracy and speed, paving the path to usable brain-controlled hearing devices. These findings suggest the need for further exploration of self-supervised speech representations in auditory neural decoding and their potential to improve our understanding of how the human brain makes attentional selections.